\documentclass[a4paper]{article}
\usepackage[noblocks]{authblk}
\usepackage[all]{xy}
\usepackage{amssymb}
\usepackage{enumitem}
\usepackage{graphicx}
\usepackage{amsmath}
\usepackage{amsfonts}
\usepackage{verbatim}
\usepackage{hyperref}
\usepackage{color}

\newcommand{\true}{\ensuremath{\mathbf{t}}}
\newcommand{\false}{\ensuremath{\mathbf{f}}}

\newcommand{\eval}[1]{[\![ #1 ]\!]}

\newcommand{\bool}[0]{\ensuremath{\mathbb{B}}}

\newcommand{\streama}[0]{\ensuremath{\overline{{A}}}}

\newcommand{\squishlist}{ 
\vspace{-2mm}
   \begin{list}{$\bullet$}
    { \setlength{\itemsep}{0pt}      \setlength{\parsep}{1pt} 
      \setlength{\topsep}{8pt}       \setlength{\partopsep}{-3pt}
      \setlength{\leftmargin}{1.5em} \setlength{\labelwidth}{1em}
      \setlength{\labelsep}{0.5em} } 
      }
\newcommand{\squishlisttwo}{
\begin{list}{$\bullet$}
  { \setlength{\itemsep}{0pt}
    \setlength{\parsep}{0pt}
    \setlength{\opsep}{0pt}
    \setlength{\partopsep}{0pt}
    \setlength{\leftmargin}{2em}
    \setlength{\labelwidth}{1.5em}
    \setlength{\labelsep}{0.5em} } }
\newcommand{\squishend}{
    \end{list}  }

\begin{document}
\title{Monitoring and Intervention: Concepts and Formal Models}

\author{Kenneth Johnson}
\affil{School of Engineering, Computer and Mathematical Sciences\authorcr 
Auckland University of Technology\authorcr
Private Bag 92006, Auckland, 1142, New Zealand}
\author{John V Tucker}
\affil{Department of Computer Science,
College of Science\authorcr Swansea University\authorcr
Singleton Park, Swansea, SA2 8PP, United Kingdom}
\author{Victoria Wang}
\affil{University of Portsmouth,
Institute of Criminal Justice Studies, 
University of Portsmouth\authorcr
St George's Building, 141 High Street, Portsmouth\authorcr
PO1 2HY,United Kingdom}

\maketitle
\begin{abstract}
\noindent Our machines, products, utilities, and environments have long been monitored by embedded software systems. Our professional, commercial, social and personal lives are also subject to monitoring as they are mediated by software systems. Data on nearly everything now exists, waiting to be collected and analysed for all sorts of reasons. Given the rising tide of data we pose the questions: \textit{What is monitoring? Do diverse and disparate monitoring systems have anything in common?} We attempt answer these questions by proposing an abstract conceptual framework for studying monitoring. We argue that it captures a structure common to many different monitoring practices, and that from it detailed formal models can be derived, customised to applications. The framework formalises the idea that monitoring is a process that observes the behaviour of people and objects in a context. The entities and their behaviours are represented by abstract data types and the observable attributes by logics. Since monitoring usually has a specific purpose, we extend the framework with protocols for detecting attributes or events that require interventions and, possibly, a change in behaviour. Our theory is illustrated by a case study from criminal justice, that of electronic tagging.


\end{abstract}

\section{Introduction}\label{Introduction}

Our machines, products, utilities, and environments have long been measured and monitored by embedded software systems. Monitoring is fundamental in science and engineering, where instruments are created to observe phenomena inside and outside the laboratory. Monitoring is essential for the proper functioning of manufacturing plant in factories, and the infrastructures of energy, transport, communications and information. The rise of cyberphysical systems -- networks of sensors, activators and processors -- is transforming industry and its products \cite{Industrie4.0}.

However, monitoring is by no means confined to science and engineering. Accounting, insurance and other financial services, with equally long histories, have developed theories and methods to monitor money -- algebra began in the service of commerce \cite{Swetz1987}.  Today, our  commercial, social, professional and personal lives are mediated by software and so are also subject to monitoring. Data is available about all aspects of our everyday life Ð as individuals, or members of groups, organisations and societies Ð to be collected, analysed, and compared for all sorts of reasons. Monitoring has made surveillance and privacy an international public concern.\footnote{For example, in the UK, privacy in monitoring and surveillance has been addressed in the Anderson report to the government \cite{Anderson2015}.} 

Despite the fact that monitoring practices are ubiquitous, and are the source of data that drives the development of data science, the nature of monitoring has been neglected theoretically. We pose the questions: 
\newline
\newline
\textit{What is the nature and purpose of monitoring? Do the diverse and apparently disparate monitoring systems have anything in common?}
\newline

We attempt to answer these questions by proposing an abstract approach to monitoring that can explore the common structure of many different monitoring systems. The scope of monitoring is colossal. Monitoring examples abound in science, engineering, commerce, manufacturing, infrastructure, healthcare, management, security, and services. By reflecting on many monitoring examples, we have isolated some essential conceptual components of monitoring systems. The case study we choose to use to illustrate our theory is the remote monitoring of offenders in criminal justice jurisdictions. 

The analysis results in an abstract conceptual framework that is intended to capture and illuminate a wide spectrum of monitoring practices. From the conceptual framework more detailed formal models can be derived that are customised to particular application domains. From the conceptual framework we derive a general mathematical model of monitoring in which behaviours are modelled by streams, i.e., sequences of data indexed by time. We use the term conceptual framework because it isolates concepts that can be formalised in a number of ways. For example, streams come in several forms: total or partial maps, defined for finite or infinite discrete or continuous time intervals. 

It must be emphasised that ours is a theoretical investigation of ideas about \textit{monitoring systems}, intended to raise and make precise general questions, to enable comparisons to be made, and to develop useful classifications of monitoring systems.  Hopefully, our formal models and methods will help the analysis of both technical and sociological issues to do with monitoring. In launching these ideas we have chosen to keep our formal techniques simple, assuming little more than the basic algebraic definitions of abstract data types: see \cite{Meinke1992,SpecADT1996} for the basic theory.

The conceptual framework formalises the idea that monitoring is a process that observes properties of the behaviour of people and objects in a \textit{context}. A context is characterised by \textit{entities} and their \textit{characteristics} and  \textit{behaviours}. Monitoring a context begins by choosing \textit{attributes} to be observed. The contexts are represented by abstract data types and the observation attributes by logical languages.  Our conception of monitoring is based upon the following principle: \textit{monitoring is confined to the collection, evaluation and recording of observational data; the outputs of a monitoring system are simply records}.  Thus, a key idea in our analysis is that it is only concerned with data, and we separate the acquisition of the monitoring data from its subsequent use. This separation enables the framework to capture commonalities of diverse domains, and increases the generality of the analysis. 

However, the subsequent use of the recorded observations is of importance, since monitoring usually has a specific purpose. We extend the monitoring framework with protocols for detecting special attributes that require attention. The records are inspected and certain attributes allowed to trigger actions that transform the record and can, in turn, change the behaviour of the entity.  We call these checks and transformations  \textit{interventions}. Adding this process we define a \textit{monitoring system with interventions}. The transformed record is communicated to a system independent of the monitoring system.

To complete the description of a monitoring system, we introduce an informal notion of a \textit{monitoring infrastructure}, which obtains and sends the data representing the behaviour of entities to a monitoring system; and an \textit{intervention infrastructure} that receives the information from a monitoring system with interventions and initiates various responses and actions on the entities. 

\begin{figure}
\centering
\includegraphics[scale=0.35]{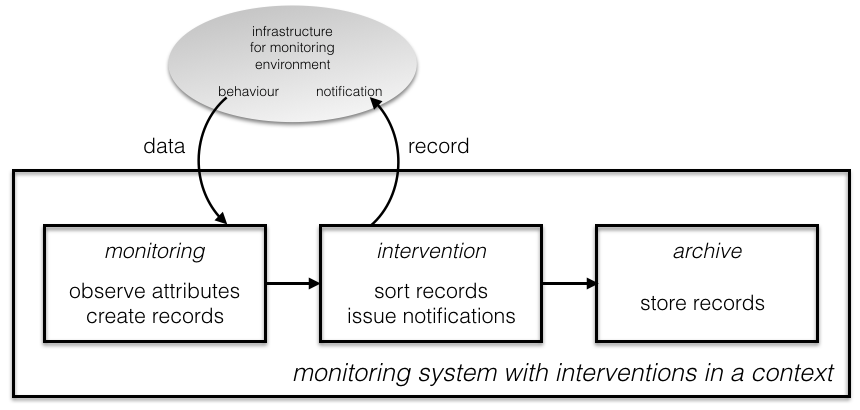}
\caption{The monitoring and intervention architecture of entities within a context}
\label{fig:framework}
\end{figure}

We assume that both the monitoring system and the interventions are composed exclusively of data, and we gather the data types into what we call the \textit{monitoring and intervention stack}. Outside the stack we allow monitoring and intervention to employ any kind of technology and practice. The flow of data is illustrated in  Figure \ref{fig:framework}. There are distinct types of intervention that reflect the purpose of monitoring. 

The paper is in three parts. In Section \ref{ConceptualFramework} we define the conceptual framework in terms of the main concepts of entity, behaviour, attribute, observation, judgement, and monitoring and interventions.  In Section \ref{BehaviourStreams} we derive a formal model from the framework by taking behaviours to be modelled by streams of data. Section \ref{SpecificationStream} summarises properties of streams in preparation for the formal specification of attributes and interventions in Section \ref{SpecificationAttributes}. In Section \ref{CriminalJustice} we consider case studies of monitoring in criminal justice to illustrate the conceptual framework and, in Section \ref{AlcoholStream}, the stream model. Finally, in Section \ref{ConcludingRemarks}, we make some remarks on future developments.


\section{Conceptual Framework of Monitoring and Interventions}\label{ConceptualFramework}

What do diverse examples of monitoring  have in common?  When we think of monitoring, we think of people and objects in the physical and virtual environments. What is monitored is some data about the behaviour of people and objects in some context and for some purpose. The monitoring of a context is characterised by specifying particular attributes that can be observed and recorded; records are outputs of the monitoring system. The purpose of a monitoring system is reflected in the application of tests that sort the observational records. If certain attributes are recognised and judged significant then notifications are issued. These notifications lead to actions that change the behaviour of the people or objects. 

\subsection{Monitoring}\label{Monitoring}

\noindent \textbf{Context: Entities, Characteristics and Behaviour.} Monitoring takes place in a context.  A \emph{context} is composed of \emph{entities}. The entities have \emph{characteristics} that define properties that are relevant to the entities in the context. Entities have \emph{behaviours} that can be observed and are monitored. Behaviour of an entity depends upon the characteristics of the entity. For example, a change in a characteristic implies a change in behaviour: characteristics are a parameter of behaviour.

A context can be specified by three sets and a function. Let $Entity$ be the set of entities. Let $Characteristics$ be the set of characteristics based upon information about the  entities in the context. Let $Behaviour$ be the set of all possible behaviours of the entities in the context. Define the behaviour mapping
\begin{align}
\eval{-,-} : Entity \times Characteristics \to Behaviour.
\label{eq:behave}
\end{align}
\noindent The map defines the semantics of the entities in the monitoring context.
\newline
\newline
\noindent \textbf{Observation: Attributes and Judgements.} Behaviours have \textit{attributes} that can be observed. Observation takes the form of making a query about, or testing, the behaviour of an entity for the presence of an attribute and making an evaluation. The evaluation gives rise to a \textit{judgement}. 

In some cases, the attribute will be a property that either holds or does not hold, and the evaluation will be a boolean judgement of $true$ or $false$. In other cases the attribute may be a property that needs assessment using a range of values on a scale; the evaluation will be a grading, measure, or probability. For example, physical measurements with error margins can give judgements that are bands of numerical values. A judgement may be a qualitative assessment based upon bands labelled metaphorically, such as the commonly used traffic light signifiers 
\begin{center}
$\{green, yellow, red\}$.
\end{center}
Let $Attribute$ be a set of attributes of behaviours. Let $Judgement$ be a set of possible evaluations of the attributes. 

We define observation in terms of attributes that can be observed.
\newline
\newline
\noindent \textit{Uniform Observation}. Consider the observation mapping that examines all behaviours with the same attribute:
\begin{align}
Obs : Attributes \times Behaviours \to Judgement\label{eq:uniformobs}
\end{align}
such that for attribute $P$ and behaviour $\sigma$,
\begin{center}
$Obs(P,\sigma) =  \textrm{a measure of the extent or degree that $P$ is a property of $\sigma$.}$
\end{center}
\noindent \textit{Individual Observation}. An attribute may vary according to the entity and its characteristics.  Thus, we may have a mapping that varies the attribute:
\begin{align}
P: Entity \times Characteristics \to Attributes\label{eq:attributeselection},
\end{align}
allowing the dependency $P(e,\chi)$.
In this case the observation mapping can take the form:
\begin{align}
Obs : Entity \times Characteristics \times Behaviours \to Judgement\label{eq:individualobs}
\end{align}
such that an entity $e$ with characteristics $\chi$, it check attribute $P(e,\chi)$ on behaviour $\sigma$:
\begin{center}
$Obs(P(e,\chi),\sigma) = \text{the extent or degree that $P(e,\chi)$ is a property of $\sigma$.}$
\end{center}

\noindent \textbf{Records.} The purpose of monitoring is to make an \textit{observation} and a \textit{record} of the observation. First, we define the form of a record: suppose $e \in Entity$, with characteristics $\chi \in Charactersitics$, has a behaviour that is tested for attribute $P \in Attribute$ with the result of the evaluation being a judgement $j \in Judgement$. Then the observation is recorded as
\begin{center} 
$(e,\chi,P, j)$.
\end{center}
\noindent Thus, the the set of all possible records of the context is
\begin{align}
Record = Entity \times Characteristics \times Attribute \times Judgement.\label{eq:record}
\end{align}

\noindent \textbf{Monitoring.} 
\noindent Observable attributes enable entities and their behaviours to be monitored. 
\newline
\newline
\noindent \textit{Uniform Monitoring}. We define the mapping 
\begin{align} 
Monitor : Entity \times Characteristics \times Attribute  \to Record\label{eq:uniformmonitormap}
\end{align}
\noindent such that
\begin{align}
Monitor(e,\chi,P) = (e,\chi,P,Obs(P,\eval{e, \chi})).\label{eq:uniformmonitormapequation}
\end{align}
The record $Monitor(e,\chi,P)$ is the output of the monitoring system.
\newline
\newline
\noindent \textit{Individual Monitoring}. If the attribute $P$ depends upon entity $e$ and characteristic $\chi$ then we see $P(e,\chi)$ in place of $P$. The monitor function now becomes
\begin{align} 
Monitor : Entity \times Characteristics  \to Record\label{eq:individualmonitormap}
\end{align}
\noindent defined by
\begin{align}
Monitor(e,\chi) = (e,\chi,P(e,\chi),Obs(P(e,\chi),\eval{e, \chi})).\label{eq:individualmonitormapequation}
\end{align}
\noindent This is an important case as we will see shortly.

\subsection{Interventions}\label{Interventions}

The purpose of monitoring is to observe attributes of interest and to record them. It is concerned with the assembly and evaluation of data. To use the monitoring data, properties must be recognised in the records, noted and communicated to infrastructures \textit{outside} the monitoring system. These communications we call \textit{notifications}. The infrastructures receiving notifications are called intervention infrastructures. The notification may initiate a response, which may be a series of physical or virtual actions that change an entity's characteristics and its behaviour. We formalise these stages as follows.
\newline
\newline
\noindent \textbf{Triggers.}  A attribute of behaviour is inspected and recognised in a monitoring record by a \textit{trigger condition}. A trigger condition $tc$ is mapping of the form 
\begin{align}
tc : Judgement \to Boolean\label{eq:trigger},
\end{align}
taking as input an evaluation from $Judgement$ and returning a Boolean value to note recognition or not. If judgements are booleans then the condition might simply detect an attribute is true or false, and $tc$ might be the identity function. If judgements are numbers then the condition might detect that a numerical threshold value is passed. 
Let $Trigger$ be the set of all trigger conditions.
\newline
\newline
\noindent \textbf{Actions.} An inspection of the record leads to an \textit{action} that is a notification. An action $act  \in Action$ is a mapping of the form
\begin{align}
act : Characteristics \to Characteristics\label{eq:act}
\end{align}
\noindent which changes or updates the characteristics of an entity. 

Let $null : Characteristics \to Characteristics$ be the identity map, the action that does nothing.
\newline
\newline
\noindent \textbf{Interventions.}
Triggers decide what should happen; actions change characteristics that makes things happen. An \emph{intervention}  is a rule of the form ``if $tc$ then do $act$ else do $null$"; we write this 
\begin{center}
$int : tc \to act$.
\end{center}
Let $Intervention = Trigger \times Action$ be the set of interventions assigned to the context.
\newline
\newline
An intervention $tc \to act \in Intervention$ is applied to a record
$(e,\chi,P,j) \in Report$ by the mapping 
\begin{align}
Int : Record \times Intervention \to Entity \times Characteristic \label{eq:interventionmap}
\end{align}
defined by
\[
Int((e,\chi,P,j),tc \to act) = 
\begin{cases}
(e,act(\chi)) & tc(j)\\
(e,null(\chi)) & \lnot tc(j).
\end{cases}
\]

\subsection{Architecture of the Framework: Monitoring and Intervention Stack}\label{Architecture}

The conceptual framework is intended to be a high-level architecture that isolates and names conceptual components. We have partially modelled the architecture mathematically to make precise the functions performed by the components. Essentially, there are three components to the framework: (i) a context, (ii) a monitoring system for the context, and (iii) an intervention system for the monitoring system. 
\newline
\newline
\noindent\textit{Interaction of Components. } In summary, the components interact as follows. 
In a context, given an entity $e$ with characteristic $\chi$, the behaviour in the context is $\eval{e,\chi}$. 

Choosing an attribute $P$ we observe and judge the entity's behaviour and produce a record $Monitor(e, \chi, P)$. 

The judgement in this record is tested by an intervention $int$ and we get the new characteristic $Int(Monitor(e, \chi, P), int)$. 

This may lead to a new behaviour $\eval{e,Int(Monitor(e, \chi, P), int)}$.
\newline
\newline
\noindent\textit{Monitoring and Intervention Stack. } 
The components can be combined and modelled using the theory of abstract data types. Focussing on the input-output functions we note the three many sorted algebras:

\begin{center}
$Context =  (Entity, Characteristics, Behaviour | 
\newline \eval{-,-} : Entity \times Characteristics \to Behaviour)$,
\label{eq:context}
\end{center}

\begin{center}
$Monitor =  (Context, Attributes, Judgement, Record |  \newline Obs : Attributes \times Behaviours \to Judgement, \newline Monitor : Entity \times Characteristics \times Attribute   \to Record)$,
\label{eq:monitor}
\end{center}

\begin{center}
$Intervention =  (Context, Judgement, Record, Intervention | \newline Int : Record \times Intervention \to Entity \times Characteristic).$
\label{eq:monitor}
\end{center}

\noindent  Note the algebras are built upon each other and we will call the collection the \textit{monitoring and intervention stack}. Underneath this stack of three monitoring structures are algebras from which models for $Context$ etc. can be made. We suppose these algebras are combined to make a single algebra, containing what may be needed to build the stack, that we will call the \textit{platform algebra} for the monitoring and intervention stack.

The stack and platform can be represented more formally. Strictly speaking, we could express the framework in terms of signatures and simple equations. However, at this early stage of analysing monitoring, it would be premature to apply abstract data type theory -- e.g., signatures, equational specifications, loose, initial and final semantics, term rewriting etc.  We prefer to introduce monitoring without these prerequisites, and develop a larger catalogue of exemplars; and also to leave open a choice from a wider selection of formalisation methods. 


\section{Modelling Behaviour as Streams}\label{BehaviourStreams}

The general conceptual framework for monitoring in Section \ref{ConceptualFramework} can now be refined in a number of ways to create a range of general mathematical models of monitoring and intervention. Clearly, there are different ways to model $Behaviour$, different logics to formalise $Attributes$, and different models of computation to analyse monitoring. In this section we focus on behaviour and the idea that it is dynamic, i.e., behaviour changes in time. We use the general framework as a template to build a class of monitoring models for contexts where behaviour is modelled by data streams. 

We model a behaviour of an entity changing in time by a \textit{stream} of data from $A$ in time $T$:
\begin{center}
$\ldots, a(t),\ldots \in A$ for $t \in T$.
\end{center}
There are a number of choices for stream behaviour:

(i) time can be continuous or discrete;

(ii) streams can be finite, infinite or both;

(iii) streams can be always well-defined or partial.

\noindent We choose discrete time streams that are infinite and always well-defined.  

To help contrast the conceptual framework with the detailed mathematical model, we revert to conventionally concise notations -- e.g., $Entity$ becomes $E$ etc.


\subsection{Monitoring Streams}\label{MonitoringStreams}

\textbf{Time.}  We model time by a set $T$ of data that mark points in time. Commonly, in modelling physical behaviour, time is assumed to be continuous and is represented by a subset of the real numbers $\mathbb{R}$, or rational numbers $\mathbb{Q}$. In modelling computational behaviour, time is assumed to be discrete and is represented by a subset of integers $\mathbb{Z}$ or natural numbers $\mathbb{N}$. So, later, we assume that time is discrete, and take  $T = \{0,1,2,\ldots\}$ examples. 
\newline
\newline
\noindent \textbf{Behaviour.} Suppose the behaviour of entities takes place in time, represented by the set $T$.  Behaviour is characterised by some data from a set $A$ -- typically measurements, text, images or audio. Thus, we define a stream of data as a \textit{total function} $a : T \to A$ mapping time
points in $T$ to data elements in $A$. The space of all behaviours is the set $[T \to A]$ of \emph{streams}. In the case of our monitoring examples, the streams are sequences in discrete time of the form:
\begin{center}
$a(0),a(1),a(2), \ldots, a(t),\ldots \in A$ for $t \in T$.
\end{center}

\noindent \textbf{Contexts: Entities, Characteristics and Behaviour.}
Let $E$ be the set of entities and let $C$ be the set of characteristics. We define the behaviour of an entity as a stream of data, generated by the behaviour map, after equation (\ref{eq:behave}).
\begin{align}
 \eval{-,-} : E \times C \to [T \to A]\label{eq:streamsemantics} 
\end{align}
\noindent such that for entity $e \in E$, with characteristics $\chi \in C$, at time $t \in T$:
\begin{center}
$\eval{e,\chi}(t) =$ data characterising behaviour of entity $e$ with characteristics $\chi$ at time $t$.
\end{center}

\noindent \textbf{Observation: Attributes and Judgements.}
Behaviours have attributes that can be observed over time. Let $Attr$ be a set of attributes of behaviours.  Since behaviour is time dependent, attributes are time dependent; indeed it is common to look for changes over, say, an interval $[t_{1}, t_{2}] \subset T$.  Let $J$ be a set of judgements; often, $J$ is a finite set.

Mathematically, we define the act of observing an entity and its characteristics, and making an evaluation, in two ways:
\newline
\newline
\noindent \textit{Uniform Observation}. 
Consider the observation mapping that examines all behaviours with the same attribute, after equation (\ref{eq:uniformobs}):
\begin{align}
Obs : Attr \times [T \to A]  \to J \label{eq:streamobs1}
\end{align}
such that for attribute $P$ and stream $a$ 
\begin{center}
$Obs(P,a) =$ a measure of the extent or degree that $P$ is a property of the stream $a \in [T \to A]$
\end{center}
\noindent \textit{Individual Observation}. Suppose an attribute may vary according to the entity and its characteristics, after equation (\ref{eq:attributeselection}) defined by 
\begin{align}
P: E \times C \to Attr.
\end{align}
In this case the observation mapping follows the form of equation (\ref{eq:individualobs})
\begin{align}
Obs : E \times C \times [T \to A] \to J\label{eq:streamobs2}
\end{align}
and given an entity $e$ with characteristics $\chi$, checks attribute $P(e,\chi)$ on behaviour $a \in [T \to A]$:
\begin{center}
$Obs(P(e,\chi),a) =$ \text{the extent or degree that $P(e,\chi)$ is a property of $a$.}
\end{center}

\noindent \textbf{Monitoring.}
Now, we implement monitoring as follows. First, following equation (\ref{eq:record}), let 
\begin{align}
R = E \times C \times Attr \times J\label{eq:streamrecord}
\end{align} 
be the set of \emph{records}. 
\newline
\newline
\noindent \textit{Uniform Observation}. Using equation (\ref{eq:uniformmonitormap}) and equation (\ref{eq:uniformmonitormapequation}), We define monitoring by the function
\begin{align}
Monitor : E \times C \times Attr  \to R\label{eq:streamuniformmonitormap}
\end{align}
\noindent such that
\begin{align}
Monitor(e,\chi, P) = (e,\chi,P,Obs(P,\eval{e,\chi})).\label{eq:streamuniformmonitormapequation}
\end{align}

\noindent \textit{Individual Monitoring}. If the attribute $P$ depends upon entity $e$ and characteristic $\chi$ then we see $P(e,\chi)$ in place of $P$. Using equation (\ref{eq:individualmonitormap}) and equation (\ref{eq:individualmonitormapequation}), the monitor function now becomes
\begin{align} 
Monitor : E \times C  \to R\label{eq:streamindividualmonitormap}
\end{align}
\noindent defined by
\begin{align}
Monitor(e,\chi) = (e,\chi,P(e,\chi),Obs(P(e,\chi),\eval{e, \chi})).\label{eq:streamindividualmonitormapequation}
\end{align}

\subsection{Interventions for streams}\label{InterventionStreams}

Following the general framework, interventions are based upon judgements, they do \textit{not} involve behaviours directly, and so they are independent of the streams; there are only changes due to substitutions.
\newline
\newline
\noindent \textbf{Triggers.} Trigger conditions accept as input a judgement value $j \in J$, obtained as a result of the observations made by the function $Obs$, and outputs a truth value. In symbols, 
\begin{align}
tc : J \to \bool\label{eq:streamtrigger}
\end{align}
We denote the set of all trigger functions as $Trig =[J \to \bool]$. If $J = \bool$ then $tc$ is the identity or negation or a constant.
\newline
\newline
\noindent \textbf{Actions.} An action function 
\begin{align}
act : C \to C\label{eq:streamact}
\end{align}
such that 
$act(\chi)$ performs an update to the information $\chi \in C$. We denote the set of all action functions by $Act$.
\newline
\newline
\noindent \textbf{Interventions.} With both of these functions, we define
an intervention of the form $tc \to act$, where $(tc, act) \in Trig \times Act$. We use triggers and action functions 
to specify the intervention that results from the observation of an entity's behaviour. Mathematically, for $Intv =Trig \times Act$, we define the function 
\begin{align}
Int : R \times Intv \to E \times C\label{eq:streamintervention} 
\end{align}
\noindent defined by
\[
Int((e,\chi,j),\tau : c \to a) = \begin{cases}
(e,act(\chi)) & tc(j)\\
(e,null(\chi)) & \lnot tc(j).
\end{cases}
\]

\section{Platforms for Stream Behaviour}\label{SpecificationStream}

The stream model of behaviour is made from data and time. We will outline the forms of some simple abstract data types for data, time, and streams that could serve as a platform algebra for the monitoring and intervention stack for the stream model.

\subsection{Specification of Streams over Data}\label{sec:streams}

\noindent \textbf{Data.} Data is never without operations and tests. Consider the set $A$ of data representing what is observable in the context. For it to be of any use we must suppose there are constants, operations and tests on $A$.  Thus, for simplicity, let the set $A$ of data be contained in a data type modelled by an algebra with one sort of data together with the Booleans $\bool$, having the form:
\begin{align}
A =  (A,\bool;c_1,\ldots,c_k,\true,\false; f_1,\ldots,f_p,r_1,\ldots,r_q,\land,\lnot),
\label{eq:datatest}
\end{align}
\noindent comprising
\squishlist
\item carriers $A$ and $\bool$;
\item  constants $c_1,\ldots,c_k$ and $\true,\false$;
\item  operations
$f_1,\ldots,f_p$ and boolean connectives $\land,\lnot$;
\item tests
$r_1,\ldots,r_q$.
\squishend

\noindent \textbf{Time.} Similarly, consider the set $T$ of time points. It, too, must belong to a data type having constants, operations and tests. Since we are assuming discrete time and that $T = \{0,1,2,\ldots\}$, to the set $T$ we add the constant $0 \in T$, the tick operation $tick:T \to T$, defined by $tick(t) = t+1$,  and equality to form an algebra of the form: 
\begin{align}
T = (T | 0, tick, =).\label{eq:time}
\end{align}
In calculating with time, other standard arithmetical operators and tests will be needed.
\newline
\newline
\noindent \textbf{Streams.} The streams of data from $A$ timed by $T$ must also belong to a data type. This leads to a \emph{stream algebra} $\streama$
that contains and \textit{expands} the algebras $A$ of data and $T$ of time with the carrier set $[T \to A]$ and a selection of operations and tests, including
the \textit{evaluation function} 
\begin{align}
eval : [T \to A] \times T \to A
\end{align} such that 
\begin{align}
eval(a,t) =  \textrm{the value $a(t)$ of the stream $a$ at time point $t$}.\label{eq:eval}
\end{align}
Constants and operators typically take the form:
\squishlist
\item
stream constants $C_1,\ldots,C_h  \in  [T \to A]$;
\item
stream operations  $F_1,\ldots,F_k$, of the form
$F_i : [T \to A]^{m_i} \to [T \to A]$, for $1 \le i \le k$; 
\item
stream operations  $F_1,\ldots,F_k$, of the form
$F_i : [T \to A]^{m_i}  \times T \to  A$, for $1 \le i \le k$; 
\squishend
and tests typically take the form
\squishlist
\item
stream tests  $R_1,\ldots,R_l$, of the form
$R_i : [T \to A]^{m_i} \to \bool$, for $1 \le i \le l$; 
\item
stream tests  $R_1,\ldots,R_l$, of the form
$R_i : [T \to A]^{m_i} \times T \to \bool$, for $1 \le i \le l$.
\squishend

\subsection{Examples of Stream Algebras}\label{Examples2}
Stream algebras are custom built for applications. For the purpose of illustration, we consider some examples of stream algebras $\streama$ over $A$ made by simply choosing operations on streams. 
In all our examples, time $T$ is fixed to be the time algebra defined in
(\ref{eq:time}). Simple examples of constants, operations and tests in $\streama$ that are useful in many situations are the pointwise liftings of the constants, operations and tests on $A$:

\squishlist
\item
For a data constant $c \in A$, stream constant $C \in  [T \to A]$ defined by $C(t) = c$ for all $t \in T$.

\item
For a  data operation $f : A^m \to A$,  stream operation  $F: [T \to A]^{m} \to [T \to A]$ defined by
\begin{center}
$F(a_1, \ldots, a_m)(t) = f(a_1(t), \ldots, a_m(t))$ for all $t \in T$. 
\end{center}

\item 
For a  boolean test  $r : A^m \to \bool$ on data,  stream test operation  $R: [T \to A]^{m} \to [T \to \bool]$ defined by 
\begin{center}
$R(a_1, \ldots, a_m)(t) = r(a_1(t), \ldots, a_m(t))$ for all $t \in T$. 
\end{center}

\squishend

\noindent The operations may return data rather than streams of data, for example: 
\squishlist
\item 
For an  operation $f : A^m \to A$ on data, the operation  $F: T \times [T \to A]^{m} \to A$ defined by 
\begin{center}
$F(t, a_1, \ldots, a_m) = f(a_1(t), \ldots, a_m(t))$. 
\end{center}
Here we  uncurry the  operator $F$ above.
\item 
For a  boolean test  $r : A^m \to \bool$ on data, stream test operation  $R_{<}: T \times [T \to A]^{m} \to \bool$ defined by 
\begin{center}
$R_{<}(t, a_1, \ldots, a_m) = (\forall s < t) [r(a_1(s), \ldots, a_m(s))]$. 
\end{center}

\item 
For a  boolean test  $r : A^m \to \bool$ on data, stream test operation  $R_{?}: T \times [T \to A]^{m} \to \bool$ defined by 
\begin{center}
$R_{?}(t, a_1, \ldots, a_m) = (\exists s < t) [r(a_1(s), \ldots, a_m(s))]$. 
\end{center}

\squishend

\noindent Some slightly complex operations include:
\newline
\squishlist

\item
The operation $shift : [T \to A]  \times T \to [T \to A]$
defined  by 
\[
shift(a,k)(t) = a(t+k).
\] 

\item
The operation  $merge: [T \to A]^{2} \to [T \to A]$ defined by
\[
merge(a, b)(t) =
\begin{cases}
a(t/2), 	& \text{$t$  is even;}\\
a((t-1)/2) & \text{$t$  is odd.}
\end{cases}
\]

\item
The operation  $insert: [T \to A] \times T \times A \to [T \to A]$ defined by
\[
insert(a,t,x)(s) =
\begin{cases}
x & t = s \\
a(s) & t \neq s.
\end{cases}
\]

\squishend

\section{Specification of Attributes and Interventions}\label{SpecificationAttributes}

In developing the theory, and in modelling examples, there are four primary components to explore: $Behaviour$, $Attributes$, $Judgements$ and $Interventions$. We have discussed formalising $Behaviour$ as streams. What can be expected of formalisations of $Attributes$ and $Interventions$ -- in general and in the stream model? If attributes express properties of streams then their formalisation depends on the basic operations included in a stream algebra (c.f., \ref{sec:streams}). 

\subsection{Three General Principles}\label{Principles}

In monitoring, attributes represent properties of the behaviours of entities that we wish to detect. Attributes are a primary component requiring judgements that determine interventions. In the high-level framework, we might expect to apply three principles: 
\newline
\newline
\textbf{Specification Principle.} \textit{Attributes are definable in a specification language. The semantics of the specification language is based on judgements.}
\newline
\newline
\textbf{Reasoning Principle.} \textit{Attributes can be transformed by means of rules for reasoning in an appropriate logic.}
\newline
\newline
\noindent There are several theoretical and practical options for logical languages with which to apply these two design principles; we will discuss some basic choices shortly. Perhaps this third principle is most fundamental:
\newline
\newline
\textbf{Computability Principle.} \textit{So we can observe and test behaviours, the attributes and the judgements of attributes are computable. In short, in the uniform case,
\begin{center}
$Obs : Attributes \times Behaviours \to Judgement$ 
\end{center}
or, in the individual case,
\begin{center}
$P: Entity \times Characteristics \to Attributes$ and 
$Obs : Entity \times Characteristics \times Behaviours \to Judgement$
\end{center}
\noindent are computable. In consequence, monitoring is computable.}
\newline
\newline
\noindent 
We turn first to the Specification and Reasoning Principle and illustrate simple choices for a formal language and logic to define and transform attributes based upon \textit{first order logic}. Although, its seems somewhat narrow in scope, since attributes are evaluated by booleans, first order logic is serviceable. Our expectations of the semantics of judgements are much wider. We expect that future case studies will introduce temporal and many valued logics -- from discrete judgements in three and $n$-valued logics to continuous judgements in fuzzy and probabilistic logics.

\subsection{Variants of First Order Logic}

Consider specifying and reasoning about attributes (and interventions) with variants of first order logic. Ultimately, the logics will be defined over the platform algebras from which the monitoring stack is built. In the case of a stream model of monitoring the platform will be a stream algebra of some kind. 

Let $A$ be any $\Sigma$-algebra with signature $\Sigma$ containing symbols $r_1,\ldots,r_m$ for tests that form the \emph{atomic predicates} of a language based on $\Sigma$, denoted $Pred(\Sigma)$.  Let $T(\Sigma, Z)$ be the $\Sigma$-algebra of terms.
\newline
\newline
\noindent \textit{Quantifier-Free First Order Formulae}. 
The set $QF(\Sigma,Z)$ of quantifier-free first order formulae over the signature $\Sigma$ with variables in $Z$ is defined inductively by the rules
\begin{align}
\alpha ::=
t_1 = t_2 \mid&
r_1(t_1,\ldots,t_{n_1}) 
\mid \cdots \mid
r_m(t_1,\ldots,t_{n_m})
\mid\nonumber\\
\lnot \alpha_1 
\mid& \alpha_1 \land \alpha_2
\mid \alpha_1 \lor \alpha_2,\label{eq:qffof}
\end{align}
where the $t_{i_j}$'s are terms in $T(\Sigma, Z)$,
and 
$\alpha_1$ and $\alpha_2$ are quantifier-free first order formulae.
The semantics is defined on any $\Sigma$ algebra $A$ and is derived from term evaluation.
\newline
\newline
\noindent \textit{First Order Formulae}. 
We extend the set of quantifier-free formulae defined by  
(\ref{eq:qffof}) to include the existential $\exists$ 
and universal $\forall$ symbols
to define the set $F(\Sigma,Z)$ of first order formulae as follows:
\begin{align}
\alpha ::=
t_1 = t_2 \mid&
r_1(t_1,\ldots,t_{n_1}) 
\mid \cdots \mid
r_m(t_1,\ldots,t_{n_m})
\mid\nonumber\\
\lnot \alpha_1 
\mid& \alpha_1 \land \alpha_2
\mid \alpha_1 \lor \alpha_2 
\mid \exists z : \alpha_1 \mid \forall z : \alpha_1 \label{eq:fof}
\end{align}
for $\Sigma$-terms $t_{i_j}$, atomic propositions $r_1,\ldots,r_m$ and formulae $\alpha_1,\alpha_2 \in F(\Sigma,Z)$ and variable $z \in Z$. The semantics is defined on any $\Sigma$ algebra $A$ in the usual way.
\newline
\newline
\noindent \textit{Weak Second Order Formulae}.  Let $A^{\star}$ be an $\Sigma^{\star}$-algebra that adds all finite sequences over $A$ to the algebra $A$ along with appropriate operations and tests. Defining quantifier-free and first order formulae over $\Sigma^{\star}$ yields weak second order languages $QF(\Sigma^{\star},Z)$ and $QF(\Sigma^{\star},Z)$ that are useful in working with computability and examples \cite{TuckerZucker2000,TuckerZucker2015}.

In the case of a stream model of monitoring, the logics are defined by over a stream algebra of some kind, acting as a platform algebra for the monitoring stack.

\subsection{Computability}
Consider the computability of attributes (and interventions). As with the logics, computability will need to be defined over the platform algebras from which the monitoring stack is built. 

There are different theoretical approaches to computability on abstract data types, and hence to applying the third design principle on the computability of monitoring \cite{TuckerZucker2006}. There are \textit{abstract computability theories} based on only on abstract data types \cite{TuckerZucker2000,TuckerZucker2015}. There are \textit{concrete computability theories} based on making finite representations of the abstract data types \cite{StoltenbergTucker1995,StoltenbergTucker1999}. Here we choose to use an abstract model of computation to make occasional remarks on computability: the imperative programming model \textbf{while} \textit{programs with arrays}. These are programs are built form assignments, sequencing, conditionals and iteration augmented by finite unbounded arrays. The model has universal computable functions for which the arrays are necessary. The sets and functions computable by \textbf{while} programs with arrays are the subject of a robust a Generalised Church-Turing Thesis for algorithms based upon abstract data types. The \textbf{while} programs with arrays are the subject of \cite{TuckerZucker2000,TuckerZucker2015}.

In the case of a stream model of monitoring, the computability is defined by \textbf{while} programs with arrays over some form of stream algebra, acting as a platform algebra for the monitoring stack.

In both abstract and concrete models of computability, we expect to be able to prove properties of the following form.
\newline
\newline
\noindent\textbf{Lemma.} Attributes defined by quantifier free formulae over signature $\Sigma$ are computable on the $\Sigma$-algebra $A$.
\newline
\newline
However, not all the usual operators on streams can be taken to be computable. Equality offers standard examples of non-computable operations: if $A$ has more than one element, the non-equality stream test $\neq : [T \to A]^{2} \to \bool$ defined by
\begin{center}
$a \neq b$ if, and only if, $(\exists t \in T)[a(t) \neq b(t)]$ 
\end{center}
is only semicomputable since we can search for a time where the streams differ; but the equality stream test $=: [T \to A]^{2} \to \bool$ defined by
\begin{center}
$a = b$ if, and only if, $(\forall t \in T)[a(t) = b(t)]$. 
\end{center}
is not semicomputable.
Computability is substantial topic and whilst some methodologies could be discussed further for the general framework, it would require considerable preparations to tackle stream models of monitoring \cite{TuckerZucker2011,TuckerZucker2014}.

\subsection{Specification of Observations}\label{SpecificationObservations}

Suppose a  monitoring system makes finitely many judgements,
\begin{center}
$Judgement = \{ j_1, \ldots, j_k \}$.
\end{center}
Suppose, too, that there exists a family of conditions
\[P = (P_{j_1},\ldots,P_{j_k}) \in Attributes^{k}\]
each of which is a Boolean-valued function. 
The different conditions lead to different judgements. It is simplest if the judgements and their predicates are disjoint and cover their domain. The $k$ predicates should be computable.
\newline
\newline
\noindent\textit{Uniform Observation.} Using the conceptual framework, we define the uniform observation function for a family of predicates $P \in Attribute^{k}$ and $\sigma \in Behaviour$
\begin{align*}
Obs(P,\sigma) = 
\begin{cases}
j_1 & \textrm{if $P_{j_1}(\sigma)$}\\
\ldots\\
j_k & \textrm{if $P_{j_k}(\sigma).$}\\
\end{cases}
\end{align*}

\noindent \textit{Individual Observation.}
Now suppose that there exists a family of mappings to conditions
\begin{align}
P = (P_{j_1},\ldots,P_{j_k}): Entity \times Characteristics \to Attributes^{k}\label{k-attributeselection}
\end{align} 
each of which is a Boolean-valued function. 
We can define the individual observation function for $e \in Entity$ and $\chi \in Characteristics$
\begin{align*}
Obs(e,\chi,\sigma) = 
\begin{cases}
j_1 & \textrm{if $P_{j_1}(e,\chi)(\sigma)$}\\
\ldots\\
j_k & \textrm{if $P_{j_k}(e,\chi)(\sigma).$}\\
\end{cases}
\end{align*}
In both abstract and concrete models of computability, we expect to be able to prove properties of the following form.
\newline
\newline
\noindent\textbf{Lemma.} If the family of predicates is computable then $Obs$ is computable. In consequence, monitoring is computable. 
\newline
\newline

\subsection{Specification of Interventions}\label{SpecificationInterventions}

The interventions each have the form 

\begin{center}
boolean-valued trigger condition on judgements 

$\Longrightarrow$ 

transformative action on characteristics
\end{center}

The discussion on how to specify, reason and compute with these components is similar to that of attributes in the previous section.  The triggers are predicates on judgements definable as boolean-valued formulae in a logical or programming language. The actions are functions on characteristics definable in the same or a different logical or programming language. In the simplest case, abstract data types of judgements and characteristics can provide definitions of the triggers and actions as atomic formulae and terms, built from the basic algebraic tests and operations. In designing interventions, attributes and trigger conditions can cooperate and suggest trade-offs.

For example, if monitoring is based on finitely many judgements $Judgement = \{ j_1, \ldots, j_k \}$ then the trigger is a map from $tc : Judgements \to \bool$ and issues notifications to the intervention infrastructure as specified by the finite sets
\begin{center}
$tc^{-1}(true)$ and $tc^{-1}(false)$.
\end{center} 
The diversity of monitoring contexts lead to different but more appropriate choices for terminology for these components. The relevance of the terminolgy depends on ideas on what to do in particular situations. Interventions can represent or initiate a definite course of action for the purpose of bringing about some desired result -- such as by specifying desired behaviour or merely update data in the characteristics.  

For many monitoring applications, the set of interventions can be meaningfully called a \textit{policy} for the monitoring context. Financial and commercial services come to mind, for which there is a huge collection of examples of ``terms and conditions". With a policy comes the task of detecting when customer behaviour fails to comply with the policy. The triggers detect \textit{non-compliance} and the actions attempt to bring the entity to a state of \textit{compliance}. The aim is that the operation of whole system is observed to conform to its user specification, i.e., the policy for customers.

Our reflection on examples have identified these distinct types of intervention that reflect the purpose of monitoring:

(i) access control systems; 

(ii) permission systems; 

(iii) penalty systems; 

(iv) incentive systems

(v) recommendation systems; and 

(vi) social support systems. 

\noindent Currently, we are developing a taxonomy including these types \cite{wangtucker2016}.

\section{Framework Case Study: Monitoring in Criminal Justice}\label{CriminalJustice}

In this section, we discuss one among many sources of social monitoring systems that can provide examples to illustrate our theory of monitoring: electronic tagging systems used for the remote supervision of offenders. We consider a home detention curfew monitoring system, and a remote alcohol monitoring system.  

\subsection{Electronic Monitoring}\label{ElectronicMonitoring}

The use of electronic monitoring is firmly established in criminal justice systems and is commonly referred to as \textit{electronic tagging}  \cite{Nellis1991}. In sentencing, monitoring involves the location of people in space and time, as in confinements and curfews; and the physiological states of people, as in substance abuse cases \cite{Nellis2009}. The main rationales for the use of electronic monitoring are: 

(i) detention, as in house arrest \cite{Austin2005}, 

(ii) restriction, as in exclusion orders \cite{UKGOV2000}, and 

(iii) surveillance, as in drug rehabilitation \cite{Jolin1992}. 

\noindent In cases involving confinement, the subject is required to stay in the confines of a particular address for certain hours of the day. 

There are many commercial technologies available for monitoring.\footnote{For illustration, we will cite occasionally SCRAM Continuous Alcohol Monitoring; SCRAM Remote Breath; SCRAM GPS and SCRAM House Arrest. See: http://www.scramsystems.com/index/scram/products.}  Most of these monitoring technologies involve some kind of wearable \textit{personal identification device} (PID) that is locked on to a subjectÕs wrist or ankle with tamper-proof elements to prevent removal. 

Monitoring systems can be classified by what, and how frequently, they observe and report. Two types of attribute are commonly monitored:

(i) \textit{Spatially aware systems} are able to locate the subject's PID; and  

(ii) \textit{Condition aware systems} are able to evaluate the subject's physical condition. 

Two types of recording are commonly made by monitoring systems (\cite{Ward1998}): 

(iii)\textit{Passive systems} in which there is periodic or occasional contact with a subject's PID (e.g., \cite{CROW2002}); 

(iv) \textit{Active systems} in which there is continuous contact with a subject's PID (e.g., \cite{Rondinelli1997});

An established technology is to fit a home with sensing devices that registers the presence of the PID.\footnote{There can also be a miniature video camera that enables officials to see the PIDÕs location and activities \cite{FAB2000}}. When this PID is out of range, the monitoring service provider will be notified. Typically, if a failure occurs compliance staff will make an intervention, say, by attempting to contact the subject for an explanation. If the explanation is not acceptable, the subject is considered to have violated his/her order. If the violation is significant, the subject could be returned to court and re-sentenced or, in the case of an Home Detention Curfew, he/she will be sent back to prison.

More recently, GPS has been used as the technology that is suited to track a wider range of locations. Of course, GPS can also be grouped into two categories \cite{DEAK2012}: 

(i) \textit{active}: \textit{continuous} collection and \textit{continuous} communication of data; 

(ii) \textit{passive}: \textit{continuous} collection and \textit{periodic} communication of data. 

\noindent To elaborate, active GPS trackers allow the viewing of tracking data in real-time; whilst passive GPS trackers store information, e.g., a subjectÕs movement during a period of time, which can then be transferred to a computer through a batch download.\footnote{See: https://www.google.com/patents/US6014080.}  Restrictions on location and time are enforced through an alert that is triggered if the subject goes into prohibited areas. A subject's proximity to other individuals can also be regulated if those individuals also wear GPS devices, or are regularly informed of the subject's location \cite{Black2003}. 

Mobile condition aware tracking devices, such as breathalyzers, are available to track a person's location and monitor biochemical characteristics, e.g., blood-alcohol level \cite{Burns2012}.\footnote{See: https://www.google.com/patents/US7341693.} Currently, a breathalyzer is built into an alcohol monitoring system. Some of the more advanced versions of alcohol monitoring systems not only detect alcohol, but also record the face and location of the test taker.\footnote{See: http://www.scramsystems.com/index/scram/scram-remote-breath.} 

Technologies are developing, though not without ethical issues. In some special cases, miniature tracking devices can be implanted beneath the subjectÕs skin to track his/her location and monitor his/her physiological signs \cite{Black2003}. In 2002, in the UK, there had been indications that the government may consider the use of surgically implanted devices for convicted paedophiles \cite{Bright2002}. Six years later, in 2008, the UK government was planning to implant miniature tracking devices under the skin of thousands of offenders, as part of an expansion of the electronic tagging scheme, to create more space in British prisons \cite{BRADY2008}.    Ethical implications of implanting electronic tagging devices need to be considered seriously before any such developments actually take place \cite{Foster2008}.  Though recently in the UK, a paedophile has been made to wear a GPS tracker for life \cite{Russell2014}.

\subsection{Framework: Tagging}

Tagging systems have much in common -- a fact demonstrated by the following classification using our general framework in Section \ref{ConceptualFramework}. Current electronic tagging systems used in criminal justice are monitoring systems of the following general kind:   
\newline
\squishlist 
\item \emph{Context:} Remote supervision of an offender under court sentence.
\item \emph{Entity:}  The offender's PID or tagging device.
\item \emph{Identifier:} The PIDÕs identifier.
\item \emph{Characteristics:} The conditions of the sentence of the court, e.g., location, time and personal conditions.
\item \emph{Behaviours:} The movements and activities of the offender.
\item \emph{Attributes:}  The measurements of location, time and personal conditions.
\item \emph{Record:} OffenderÕs PIDÕs identifier, conditions of the sentence, evaluation of compliance.
\item \emph{Intervention:} The triggering condition tests the evaluation of compliance with the court sentences and the action notifies intervention infrastructures.
\squishend
  
In the criminal justice system, the court can sentence an offender to conform to a range of constraints on their activities (recalling our overview in Section \ref{CriminalJustice}). We will now refine this general description, customising it to particular court sentences and making formal models based on streams and logical formulae. 

To adapt the above to more closely describe a home detention curfew monitoring system (either passive and active) only a few changes are needed:
\newline
\squishlist 
\item \emph{Context:} Remote supervision of an offender in house arrest.
\item \emph{Entity:}  The offenderÕs PID or tagging device.
\item \emph{Identifier:} The PIDÕs identifier.
\item \emph{Characteristics:} The conditions of the house arrest, e.g., remaining in doors for fixed periods 19.00-07.00 of each day.
\item \emph{Behaviours:} The location of the offender.
\item \emph{Attributes:}  The measurements of location and time.
\item \emph{Record:} OffenderÕs PIDÕs identifier, location and time restrictions, check on offenderÕs residence.
\item \emph{Intervention:} Offender absent from home during curfew period triggers investigation.
\squishend

\subsection{Framework: Remote Alcohol Monitoring} 

The remote alcohol monitoring system enables agencies to accurately monitor individuals' alcohol levels, as an independent measure or in combination with a restrictive curfew monitoring schedule. Remote alcohol monitoring requires a device that can 

(i) test alcohol levels in the subject; 

(ii) establish the identity of the subject taking the test; 

(iii) create a record containing the identity and alcohol measurement; and

(iv) communicate the record to the supervising service. 

\noindent The alcohol can be tested via breath or sweat using electro-chemical techniques. Identity can be established by facial recognition techniques. Communication with a supervising service can use the phone system or email. Observations of the subject can be programmed according to a schedule, regular or random, or provide on-demand tests and/or automatic re-testing. Such devices may also include location monitoring (e.g.,  Scram alcohol monitoring, or 3MÕs remote alcohol monitoring, which can be passive or active; recall \ref{ElectronicMonitoring}).
     
First, we will consider remote alcohol monitoring in terms of our monitoring framework and then describe a formal model using streams and logical formulae.

The concentration of the alcohol in the air in the lungs is directly related to the concentration of the alcohol in the blood. The ratio of breath alcohol to blood alcohol is 2100 to 1 (and called the \textit{partition ratio}), so the alcohol content of 2100 millilitres of exhaled air will be the same as for 1 millilitre of blood. This leads to blood alcohol readings expressed as a percentage of alcohol in the blood. The partition ratio can vary (between 1700 and 2400) depending upon the individual and local environmental conditions, leading to a breath analysis reporting either a higher or lower calculated blood alcohol reading.

To take one early example of a criminal justice programme, in South Dakota, USA, there is the \textit{24/7 Sobriety Program},\footnote{See: http://apps.sd.gov/atg/dui247/.}  begun in 2005, which attempts to prevent repeat \textit{driving while intoxicated} (DWI) offenders from drinking through frequent testing.\footnote{In many countries, including England and Wales, the alcohol limit for drivers is 80 milligrams of alcohol per 100 millilitres of blood, a blood-alcohol concentration (BAC) of 0.08 \%. In Scotland and other European countries the rate is lower.} The program required repeat DWI arrestees to submit to twice-daily (7 -Ð 9 a.m. and 7 -Ð 9 p.m.) breath testing as a condition of bail. Failed tests constituted a violation of bond terms and were punishable by immediate 24-hour incarceration; and missed tests led to issuing of an arrest warrant.\footnote{See: http://onlinelibrary.wiley.com/doi/10.1111/j.1360-0443.2009.02844.x/epdf}

Four testing modalities are used: (i) twice-daily breath testing for alcohol; (ii) ankle bracelets that monitor alcohol consumption continuously with daily remote electronic reporting; (iii) twice-weekly urine testing for drugs; and (iv) sweat patches for drug monitoring (worn for 7 -- 10 days and mailed in). Random drug testing was added to alcohol testing to discourage substitution. In the following example, we consider a PID of type (ii):  
\newline
\squishlist 
\item \emph{Context:} Remote supervision of an individual's alcohol intake.
\item \emph{Entity:}  The offender's breathalyzer.
\item \emph{Identifier:} Tagging breathalyzer's identifier.
\item \emph{Characteristics:} The conditions of the sentence of the court, e.g., location, time and personal conditions.
\item \emph{Behaviours:} The alcohol intake of the offender.
\item \emph{Attributes:}  The measurements of percentage of alcohol in exhaled air.
\item \emph{Record:} OffenderÕs breathalyzer identifier, blood concentration level less than 0.02\%, breathalyser measurement $B$.
\item \emph{Intervention:} The triggering condition if breathalyser measurement B lies between $0.015 < B < 0.025$.
\squishend

\section{Stream Model Case Study: Remote Alcohol Monitoring}\label{AlcoholStream}

To illustrate the use the stream models in subsection \ref{MonitoringStreams} we make a specification of alcohol monitoring. 
\newline
\newline
\noindent\textbf{Time. } Let $T = \{0,1,2,\ldots\}$ be a set of time points counting minutes. We will consider monitoring over an interval $[t_1,t_2]$.
\newline
\newline
\noindent\textbf{Behaviour. } The quality to be monitored is blood alcohol content. Let this be measured as a percentage and so let the scale for blood alcohol be $BAC = [0,100]$. Over time, a potential variation in measurement of blood alcohol content is represented by a stream
\[b(0),b(1),b(2),\ldots,b(t),\ldots\]
which is a total function
\[b : T \to BAC_{\bot}\]
where $BAC_{\bot} = BAC \cup \{\bot\}$ and for $t \in T$
\[b(t) = \textrm{\% alcohol in the blood}\] and
\[b(t) = \bot\] means there was no reading at time $t$.
Let $[T \to BAC_{\bot}]$ be the space of all possible variations in measurements.

Suppose the technology makes a measurement every $s$ minutes. Then the offender's blood alcohol content can be sampled at times 
\[t = 0, s, 2s,\ldots,ks,\ldots.\]
For example, $s = 30$ minutes.
\newline
\newline
\noindent\textbf{Context: Entities, Characteristics and Behaviour. } 
The \emph{entities} to be monitored are offenders but what are actually monitored are the PIDs attached to them. For simplicity, we identify offenders with their PIDs,  which are known by their identifiers (e.g., serial numbers). Let $Offend$ be the set of offenders to be monitored.

The \emph{characteristics} of an offender consist of data
$(\sigma,\alpha)$ defining
\begin{enumerate}
\item the offenders \textit{sentence} $\sigma$, and
\item the status of the offender's \textit{compliance}, $\alpha$.
\end{enumerate}

\noindent Suppose a sentence $\sigma$ is specified by 
\begin{enumerate}
\item[i.] start time  $t_1$ of sentence,
\item[ii.] end time $t_2$ of sentence,
\item[iii.] sample rate $s$ for testing alcohol,
\item[iv.] alcohol limit $\epsilon\%$.
\end{enumerate}
A sentence is represented by a 4-tuple $\sigma = (t_1,t_2,s,\epsilon)$. Let $Sentence$ be the set of sentences.

Suppose the status of an offender's compliance $\alpha$ is 
\begin{itemize}
\item \textit{green}: offender's content $ < (\epsilon-\delta)\%$
\item \textit{red}: offender's content $ > (\epsilon+\delta)\%$
\item \textit{amber}: offender's content $\in [\epsilon-\delta,\epsilon+\delta]\%$
\item \textit{absent}: no measurement; e.g. $b(t) = \bot$.
\end{itemize}
where $\delta$ is an error margin. Let $Status$ be the set of possible states of offenders. 

Using the framework, the set of characteristics is 
\[Sentence \times Status.\] 
Using \ref{eq:streamsemantics}, the behaviour is determined by the offender and characteristics
\[\eval{-,-,-} : Offend \times Sentence \times Status \to [T \to BAC_{\bot}]\] where
\[\eval{o,\sigma,\alpha}(t) = \textrm{ outcome of measurement of 
alcohol \% at time $t$.}\]
\newline
\newline
\noindent\textbf{Attributes and Observations. } Attributes are dependent on entities and characteristics: we are in the case of individual observation. To observe blood content over any period  $[t_1,t_2]$ with sample rate $s$, we calculate the number $N$ of tests possible
\[N = \Big\lfloor\frac{t_2-t_1}{s}\Big\rfloor\]
and compute the upper bound over the period:
\[
Test(t_1,t_2,s,b) = \max_{k=0,\ldots,N}\{b(t_1+k\cdot s).\}
\]
If any $b(t + k\cdot s) = \bot$ then $Test$ is not defined.

Compliance with the sentence $\sigma=(t_1,t_2,s,\epsilon)$
for the period $[t_1,t_2]$ is observed by the attributes
\begin{align*}
P_{green}(\sigma)(b) &\equiv Test(t_1,t_2,s,b) < \epsilon-\delta\\
P_{amber}(\sigma)(b) &\equiv Test(t_1,t_2,s,b) \in [\epsilon-\delta,\epsilon+\delta]\\
P_{red}(\sigma)(b) &\equiv Test(t_1,t_2,s,b) > \epsilon+\delta\\
P_{absent}(\sigma)(b) &\equiv Test(t_1,t_2,s,b) \text{ is undefined.}
\end{align*}
Notice that we have a family 
\[P = (P_{green},P_{amber},P_{red},P_{absent})\]
of Boolean-valued attributes, each of which is a function
of the sentence $\sigma$. The four predicates are quantifier-free formulae. The different attributes lead to different judgements, which correspond with status. Thus, let
\[Judgement = \{green, amber, red, absent\}.\]
Using the framework, and especially subsection \ref{SpecificationObservations}, we define $observation$
\begin{align*}
Obs(P,b) = 
\begin{cases}
green & \textrm{if $P_{green}(\sigma)(b)$}\\
amber & \textrm{if $P_{amber}(\sigma)(b)$}\\
red & \textrm{if $P_{red}(\sigma)(b)$}\\
absent & \textrm{if $P_{absent}(\sigma)(b)$}.\\
\end{cases}
\end{align*}
\newline
\newline
\noindent\textbf{Monitor.  } The records have the form
\[Record = Offend \times Sentence \times Status \times Attribute \times Judgement.\]
Typically, $(o,\sigma,\alpha,P(\sigma),j) \in Record$. 
Using the stream interpretation of the framework, equation \ref{eq:streamindividualmonitormapequation}, we have 
\[Monitor : Offend \times Sentence \times Status \to Record\] and for offender $o$, sentence $\sigma$, status $\alpha$, and parameterised attribute $P$, we have
\[Monitor(o,\sigma,\alpha) = (o,\sigma,\alpha,P(\sigma),Obs(P(\sigma),\eval{o,(\sigma,\alpha)})).\]
To complete our illustration of the framework, we turn to interventions.
\newline
\newline
\noindent\textbf{Interventions. }Interventions are needed when the judgement in the monitoring record are 
\textit{red}, \textit{amber} or \textit{absent}. 

The real world intervention begins with updating the status data in the monitoring record. Using the framework, we gave the trigger condition 
\[c : Judgement \to Boolean\] defined by
\[
c(j) = \begin{cases}
false & \textrm{if $j = green$},\\
true & \textrm{otherwise}.
\end{cases}
\]
The trigger condition causes change in the characteristics of the entity by means of 
\[act : Sentence \times Status \times Judgement \to Sentence \times Status\]
if $c(j) = false$ then there is no change and \[act(\sigma,\alpha,j) = (\sigma,\alpha).\]
and if $c(j) = true$ then \[act(\sigma,\alpha,j) = (\sigma,\beta)\] where $\beta = j$.
\newline
\newline
\noindent\textbf{Computability. } It is not difficult to create an appropriate a stream algebra to act as a platform for the monitoring and intervention stack for alcohol monitoring, and to show that monitoring and interventions are computable in, say, an abstract computability model such as \textbf{while-array} programs (using section \ref{SpecificationObservations}).  
\newline
\newline
\noindent\textbf{Extensions. }We have made some assumptions 
that simplify the monitoring model and its explanation. Sentences, monitoring technologies and notification protocols can be more complicated. We note some refinements that affect time. Consider the questions:
\begin{itemize}
\item How long is a sentence?
\item Does it refer to special periods of the day?
\item How frequently are measurements made?
\item How frequently  is the offender authenticated? 
\item How frequently are measurements evaluated and reported?
\end{itemize}
Answers to these questions lead to developments of the model: new data and time parameters, and new streams are introduced. To gain an impression, we reflect on the following case.

Suppose a sentence specifies that an offender must remain ``dry'' for  period of $M$ days in the sense that his or her blood alcohol content is never more than $\epsilon \%$. Suppose the offender's content can be measured every $s$ minutes, and the measurements evaluated and reported/uploaded at certain fixed times of the day. Suppose the offender is authenticated each time the data is uploaded.
New parameters are needed for scheduling old and new features. For instance, suppose the sentence is $M=100$ days and the limit is $\epsilon = 0.02\%$. Suppose the technology makes measurements every 60 minutes, but the checking of measurements, uploading data, and authentication of the offender takes place every 8 hours at 7:00,15:00 and 23:00. The components of the model presented can be expanded to accommodate these scheduling constraints and more complicated protocols for intervention devised.


\section{Concluding Remarks}\label{ConcludingRemarks}

In this paper, we have (i) given a general conceptual framework for thinking about monitoring;  (ii) used it to develop a general mathematical model of monitoring based on modelling behaviour as streams of data; and (iii) illustrated the framework and stream model with case studies from criminal justice.  This is the first in a series of papers that aims to develop a general theory of monitoring.

\subsection{Monitoring Applications}

There are many areas where monitoring is a core activity and where the theory could be developed further and applied. Our original motivation was to provide tools to analyse the role of monitoring in governmental, civic, social and personal contexts. Monitoring is a foundation upon which studies of surveillance, privacy and identity may be built.

However, an obvious area in which to study examples of monitoring is computer systems, where every form of interaction can be recorded. For example, \textit{cloud computing services} involve several interesting forms of monitoring.\footnote{The different kinds of monitoring activities in the cloud are many and have been broadly categorised in \cite{Montes2013} according to two factors: 

(i) the \emph{architectural layer} of the cloud that is being observed, and 

(ii) the \emph{purpose} for which the monitoring is being performed.  

\noindent A similar categorisation is developed  in \cite{aceto2013}.}
 In the case where computing resources are rented to customers on a pay-as-you-go basis, the customer's account with the provider is governed by a \textit{service-level agreement} (SLAs) that 
specifies the terms and conditions for the customer and
guarantees a \emph{quality of service} (QoS) to be delivered by the provider \cite{Gar2014}. Monitoring of the SLA is needed. From the provider's perspective, continuous monitoring of the data centre's hardware and software resources in real-time is important for provisioning cost-effective services. Conversely, the customer monitors the virtual services they use to measure their QoS performance. The observations gathered from monitoring helps the customer decide on changes -- e.g., to provision more or less resources -- in response to frequently changing user requirements. 

Another obvious area is that of \emph{eHealth systems}.  Examples of eHealth monitoring systems are ubiquitous in clinical environments, of course. In addition, there are newer systems that enable rapid patient access to services and information \cite{cunningham_definitions_2014,eHealth1,second2010}.\footnote{There is also medical practice management software \cite{kareo,practoray} used by clinics to organise patient profiles, appointment schedules and billing, and large-scale information systems to manage physical resources and patient care within hospitals, all of which is monitored.}  The proliferation of mobile devices, such as smart phones and smart watches, means that users outside a clinical environment can regularly monitor and record their health data. eHealth systems are used directly by patients to assist in managing  diabetes \cite{commodity,glucmodel} or record and track heart symptoms  \cite{heartrate2013,albert_wireless_2012,alivecor}.

\subsection{Technical Developments}\label{TechnicalDevelopments}

The ideas in this paper can be developed in several directions, some of which we have indicated. There are the theoretical developments of 

(a) data types of identifiers for entities in contexts; 

(b) semantic models of behaviour in time and space (e.g., process theories involving non-determinism and concurrency);

(c) languages and logics for attribute specification; 

(d) (non-boolean) data types for making judgements; 

(e) safety and liveness properties for interventions; 

(f) computability and complexity theories for monitoring; 

(g) applications to problems in surveillance such as privacy. 

\noindent We expect that several of these directions can be developed using existing theoretical models and methods. For example, abstract data type theory can be employed to study identity \cite{WangTucker2014} and the role of space in behaviours \cite{JohnsonTucker2013}. However, to make these theoretical developments we need to collect a large variety of examples of monitoring contexts. 

It is also important to clarify the relationship between monitoring and `big data'. In both the physical world and virtual world the activities of people, the properties of objects, and the performance of services, systems and machines are represented and modelled by data. With the abundance of methods of data collection varying from sensors operating at fantastically high speeds, through automatic record keeping and mining of customers' accounts, to slow manual data entry at regular, possibly long, time intervals. Nevertheless, regardless of how it is collected, \textit{the data commonly comes from a form of monitoring}.  We argue that  \textit{it is the world's appetite for monitoring that drives its desire for data}. For example, the contemporary concerns, technologies and applications of `big data' depend upon monitoring. We believe monitoring systems to be a topic with considerable potential for general theories and diverse applications.

\bibliographystyle{plain}
\bibliography{bibliography}

\end{document}